\documentclass[twocolumn,floatfix,prl,showpacs,superscriptaddress]{revtex4}
\usepackage{helvet}
\usepackage{eurosym}
\usepackage{color}
\usepackage[latin1]{inputenc}     
\usepackage{subfigure}
\usepackage{amssymb}
\usepackage{amsmath}
\usepackage{graphicx}
\usepackage{bbold}
\usepackage{natbib}

\DeclareGraphicsExtensions{%
    .png,%
    .jpg,%
    .pdf,.PDF,%
    .mps,.jpeg,.jbig2,.jb2,.JPG,.JPEG,.JBIG2,.JB2}

\newcommand{\avg}[1]{\left< #1 \right>} 

\newcommand{\rb}[1]{\left( #1 \right)}

\begin{document}
\title{Time-delayed feedback control of the Dicke-Hepp-Lieb superradiant quantum phase transition}
\author{Wassilij Kopylov}
\affiliation{Institut f\"ur Theoretische Physik,
  Technische Universit\"at Berlin,
  D-10623 Berlin,
  Germany}
\author{Clive Emary}
\affiliation{Department of Physics and Mathematics,
University of Hull,
United Kingdom}
\author{Eckehard Sch\"oll}
\author{Tobias Brandes}
\affiliation{Institut f\"ur Theoretische Physik,
  Technische Universit\"at Berlin,
  D-10623 Berlin,
  Germany}
\date{\today}

\begin{abstract}
We apply the time-delayed Pyragas control scheme to the dissipative Dicke model via a modulation of the atom-field-coupling.
The feedback creates  an infinite sequence of non-equilibrium phases with fixed points and limit cycles in the primary superradiant regime.
We analyse this Hopf bifurcation scenario as a function  of delay time and feedback strength, and determine analytical conditions for the phase boundaries.
\end{abstract}
\pacs{
      05.30.Rt, 
      37.10.Jk, 
      05.45.-a 
      }
\maketitle


Interacting quantum systems with time-dependent Hamiltonians
offer rich and exciting possibilities to study many-body physics beyond equilibrium conditions. There has be a recent surge in generating correlated non-equilibrium dynamics in a controlled way by changing the interaction parameters as a function of time, for example by periodically modulating the coupling constants, or by abruptly quenching them.
Of particular interest then is the fate of coherent quantum dynamics and phase transitions in such scenarios, and indeed intriguing phenomena have been discussed, such as coherent control of tunneling in Bose-Einstein condensates \cite{Ligetal07},
thermalization after quenches \cite{quenchexperiments},  or dynamical and excited state quantum phase transitions  \cite{Ecketal09,CCI08}.

In this Letter, we introduce another and conceptually very different option for
driving quantum systems out of equilibrium, i.e., by modulating interaction parameters via a measurement-based feedback loop.
The time-delayed Pyragas control scheme \cite{pyragas1992continuous} that we propose here has been successfully employed  in a classical context over the past twenty years,
for example, as a tool to stabilize certain orbits in
chaotic systems or networks \cite{schoell_handbook_chaos_control,stabil_periodorbin-choe,FLU10b,SCH13}. Its key idea is to feed back the difference between two signals of the same observable at different times, such that a stabilization occurs when the delay time matches an intrinsic period of the dynamical system.

Our key idea is to generate new non-equilibrium phases via Pyragas control of
the interaction between the single bosonic cavity mode and the collection of quantum two-level systems \cite{Dicke_Modell} in Dicke-Hepp-Lieb superradiance.
The superradiant transition without control, which has been observed only recently in
cold atoms within a  photonic cavity \cite{Baumann-Dicke_qpt,Esslinger-Dicke_qpt-1,Baumann-Dicke_qpt-2,Nagy-dicke_and_bose_einstein}, has an underlying semi-classical bifurcation,
which makes it an ideal candidate to
study feedback at the boundary between non-linear (classical) dynamics and
quantum many-body systems \cite{Dicke_Chaos_and_qpt}.

Control loops of the Dicke model have been studied
in the past, for example in the form of periodical modulations of the atom-field-coupling constants
\cite{Dicke-nonequilibrium_qpt-bastidas,Baumann-Dicke_qpt-2}, the level splitting modulation \cite{Phot_production_from_Vak_to_SR-Vacanti,Extracavity_radiation_from_single_qbit-Liberato}, or as Pyragas-feedback of the cavity mode alone \cite{Dicke_Rapid_convergence_time_delay-Grimsmo}.
In our model, we condition the effective coupling strength  between the cavity and the atoms of the Dicke system - in the experiment just proportional to the laser intensity \cite{Baumann-Dicke_qpt} -  on
a difference of  photon numbers emitted from the cavity at different times. We use a mean field approach and linear stability analysis in order to show  that closed loop control dramatically affects the states in the primary superradiant regime, creating a new phase
with an infinite sequence of Hopf bifurcations between stable fixed points and limit cycles.
We also derive analytical results in the form of a single transcendental equation that determined the boundaries between
the different zones in the phase diagrams.


\textit{Open Dicke model with time delayed feedback. ---} The Hamiltonian of the Dicke model

\begin{equation}\label{eq1:Dicke-Hamiltonian}
H = \omega \hat{a}^\dag \hat{a} + \omega_0 \hat{J}_z  + \frac{g(t)}{\sqrt{2j}}(\hat{a}^\dag + \hat{a})(\hat{J}_+ + \hat{J}_-)
\end{equation}
describes the interaction between a single bosonic mode (with
frequency $\omega$ and  annihilation operator $\hat{a}$) and $N$ two level
systems (with
level splitting $\omega_0$ and collective angular momentum
operators $\hat{J}_{z,\pm}$)
with total angular momentum $j=N/2$ \cite{Dicke_Chaos_and_qpt}.
Besides we set the length of the pseudo-spin $j$ to its maximum value.
We assume an interaction between the bosonic mode and the collective angular momentum with a time-dependent coupling $g(t)$ that is modulated
by a time-delayed feedback loop. Among various models for $g(t)$, the  Pyragas form \cite{pyragas1992continuous}
\begin{equation}\label{eq1:g_modulation_noninvasiv}
 g(t) = g_0 + \lambda \left(\avg{\hat{a}^\dag \hat{a}}(t - \tau) - \avg{\hat{a}^\dag \hat{a}}(t)\right)
\end{equation}
with the time-delayed feedback of the
boson number
at two different times $t$ and $t-\tau$ and feedback strength $\lambda$
turns out to lead to the richest phase diagrams.
In the pioneering experiments for the Dicke-Hepp-Lieb phase transition in open photonic cavities \cite{Baumann-Dicke_qpt},  the form
Eq. (\ref{eq1:g_modulation_noninvasiv})
would correspond to measured, average photon fluxes (proportional to the  mean cavity photon occupation number \cite{Oeztop-excitation_of_opticaly_driven_atomic_condensate,Kopylov_Counting-statistics-Dicke})  coupled back to a pump laser.
Apart from the Pyragas delay form, this scheme is in fact close to the original feedback loops used for modulating the photon counting statistics in lasers \cite{yamamoto_statistic_feedback_and_laser}.

We note that by using mean (expectation) values in
Eq. (\ref{eq1:g_modulation_noninvasiv}) instead of operators (and
additional noise terms in a stochastic master equation \cite{Wiseman_Milburn}) for the boson occupations, we already assume a mean field description that we expect to hold for $N\to \infty$ and that we formalize in the following.

\begin{figure}[t]
\includegraphics[width=\columnwidth]{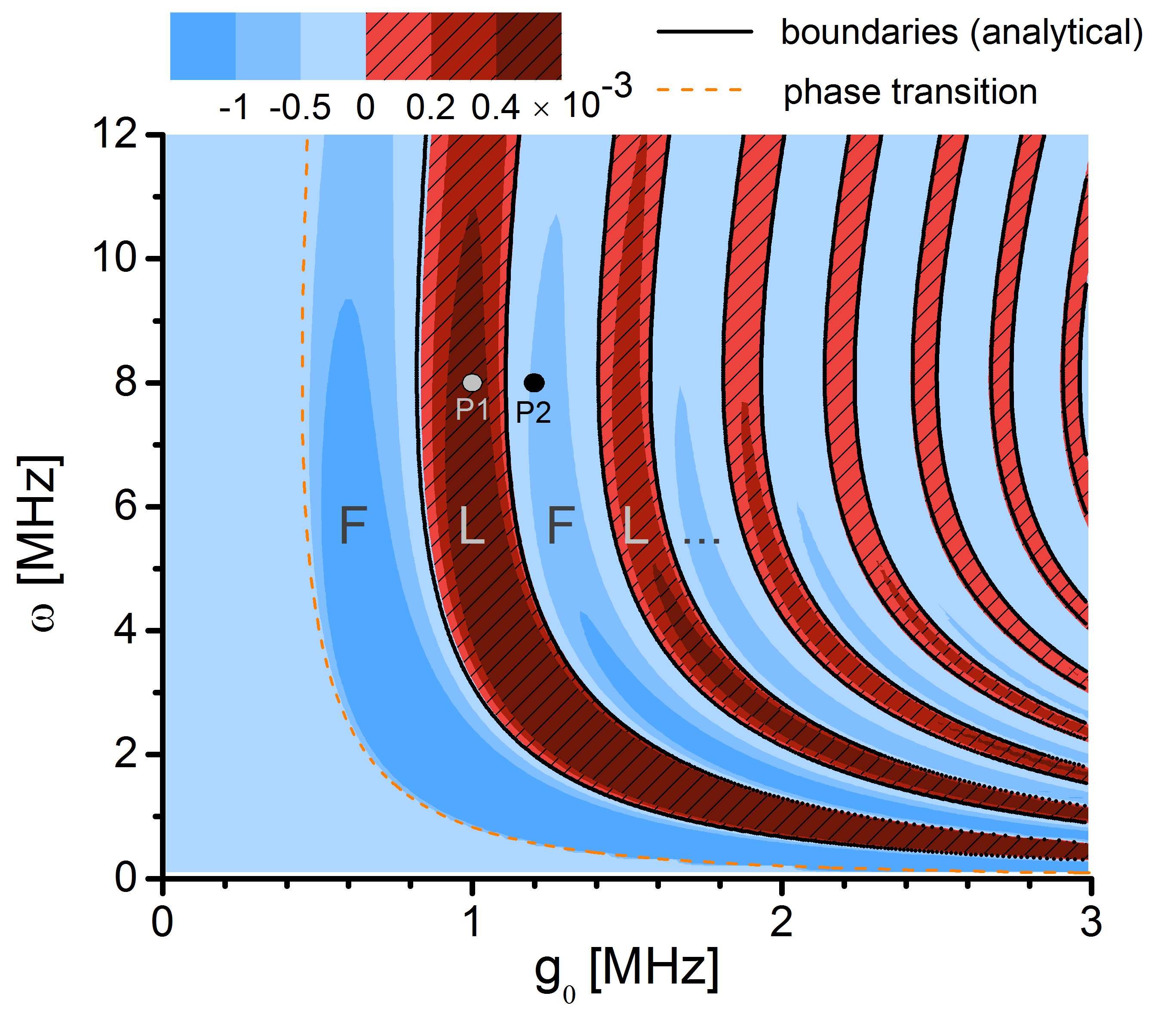}
\caption[]{\label{figure1}
Phase diagram for fixed time delay $\tau$ and feedback strength $\lambda$. The dashed (orange) line  separates the normal from the superradiant phase. The superradiant regime is split by time delayed control into
zones with stable fixed points (F) and limit cycles (L) with boundaries (black curves) determined from
Eq. \eqref{eq1:stability_end_condition_w0_is_0}. Color encodes the largest real part of the eigenvalue.
  Parameters: $\tau = 20$ $\mu$s, $\lambda = 5$ MHz and $\omega_0 = 0.05$ MHz, $\kappa  = 8.1$ MHz \cite{Bhaseen_dynamics_of_nonequilibrium_dicke_models,Baumann-Dicke_qpt}.
}
\end{figure}

\textit{Semiclassical equations. ---}
In analogy with semiclassical laser theory, phase transitions in the Dicke model for $N\to \infty$ are well described by mean-field equations for (factorized) operator expectation values \cite{Bonifacio,Bhaseen_dynamics_of_nonequilibrium_dicke_models,Kopylov_Counting-statistics-Dicke}
which we denoted by the corresponding symbols without hat.
Splitting $a$ and $J_\pm$ into real and imaginary parts, $a = x + i y$ and $J_\pm = J_x \pm i J_y$, these equations read
\begin{align}\label{eq1:open_dicke_semiclass_eq_alternativ}
\dot{x} &= - \kappa x + \omega y ,\quad
\dot{y} = - \kappa y - \omega  x - 2 \frac{g(t)}{\sqrt{2j}} J_x , \\
\dot{J}_x&= - \omega_0 J_y , \quad
\dot{J}_y = \omega_0  J_x - 4 \frac{g(t)}{\sqrt{2j}} \cdot x \cdot J_z ,\notag\\
\dot{J}_z &= 4 \frac{g(t)}{\sqrt{2j}} \cdot x \cdot J_y , \notag
\end{align}
where $\kappa$ is decay rate of the bosonic mode and where the coupling
$g(t)$ takes the form
$ g(t) = g_0 + \lambda \rb{x_\tau^2 - x^2 + y_\tau^2 - y^2}$ with the shorthand $f_\tau \equiv f(t - \tau)$.
Note that  the
angular momentum
is a conserved quantity 
even for time dependent $g(t)$, and the time development therefore  takes place on the surface of a Bloch sphere with the radius $N/2$.
For zero time-delay $\tau = 0$, i.e. without feedback, the phase diagram is well
known \cite{Bhaseen_dynamics_of_nonequilibrium_dicke_models}.
For $g < g_c\equiv \sqrt{{\omega_0 (\kappa^2 + \omega^2)}/{4  \omega}}$, a stable normal phase solution
corresponds to fixed point
$J_x^0 = J_y^0 = x^0=y^0 = 0, J_z^0 = - {N}/{2}$,
whereas
$ J_x^0 = \pm \sqrt{\frac{N^2}{4} - {J^0_z}^2}$, $J_y^0 = 0$, $J_z^0 = \frac{-N\omega_0 (\kappa^2 + \omega^2)}{8 g_0^2 \omega}$ with
$ x^0 = -J_x^0\frac{2 {g_0}\omega}{\sqrt{N}({\kappa^2 + \omega^2 })}$ , $y^0 = \frac{\kappa x_0}{\omega}
$
corresponds to the stable superradiant phase that
exists only if $g \geq g_c$.

\textit{Stability and feedback. ---}
To find out how the time-delayed feedback affects the stability of the system, we linearize Eqs.~\eqref{eq1:open_dicke_semiclass_eq_alternativ} around the fixed points (these do not depend upon $\tau$ since the feedback  Eq.~(\ref{eq1:g_modulation_noninvasiv}) vanishes in the steady state).
Using the usual procedure \cite{HOE05}
the linearized equations read
$\delta\mathbf{v}\,'(t)= \mathbf{B} \cdot  \delta\mathbf{v} (t) + \mathbf{A} \cdot \delta\mathbf{v} (t - \tau)$
with $\delta\mathbf{v} = (\delta J_x, \delta J_y, \delta x,\delta y)^T$ describing the deviation from the fixed point and
\begin{widetext}
\begin{align}
\mathbf{A} &= \left(
            \begin{array}{cccc}
             0 & 0 & 0 & 0 \\
             0 & 0 & -8 J_z^0 {x^0}^2 \lambda  & -8 J_z^0 {x^0} {y^0} \lambda  \\
             0 & 0 & 0 & 0 \\
             0 & 0 & -4 J_x^0 {x^0} \lambda  & -4 J_x^0 {y^0} \lambda  \\
            \end{array}
            \right) ,
\quad
\mathbf{B} &= \left(
            \begin{array}{cccc}
             0 & -\omega_0 & 0 & 0 \\
             \frac{4 g_0 J_x^0 {x^0}}{J_z^0}+\omega_0 & 0 & 8 J_z^0 {x^0}^2 \lambda -4 g_0 J_z^0 & 8 J_z^0 {x^0} {y^0} \lambda  \\
             0 & 0 & -\kappa & \omega  \\
             -2 g_0 & 0 & 4 J_x^0 {x^0} \lambda -\omega  & 4 J_x^0 {y^0} \lambda -\kappa \\
            \end{array}
            \right).
\end{align}
\end{widetext}

\begin{figure}[h!]
\centerline{\includegraphics[width=0.5\columnwidth]{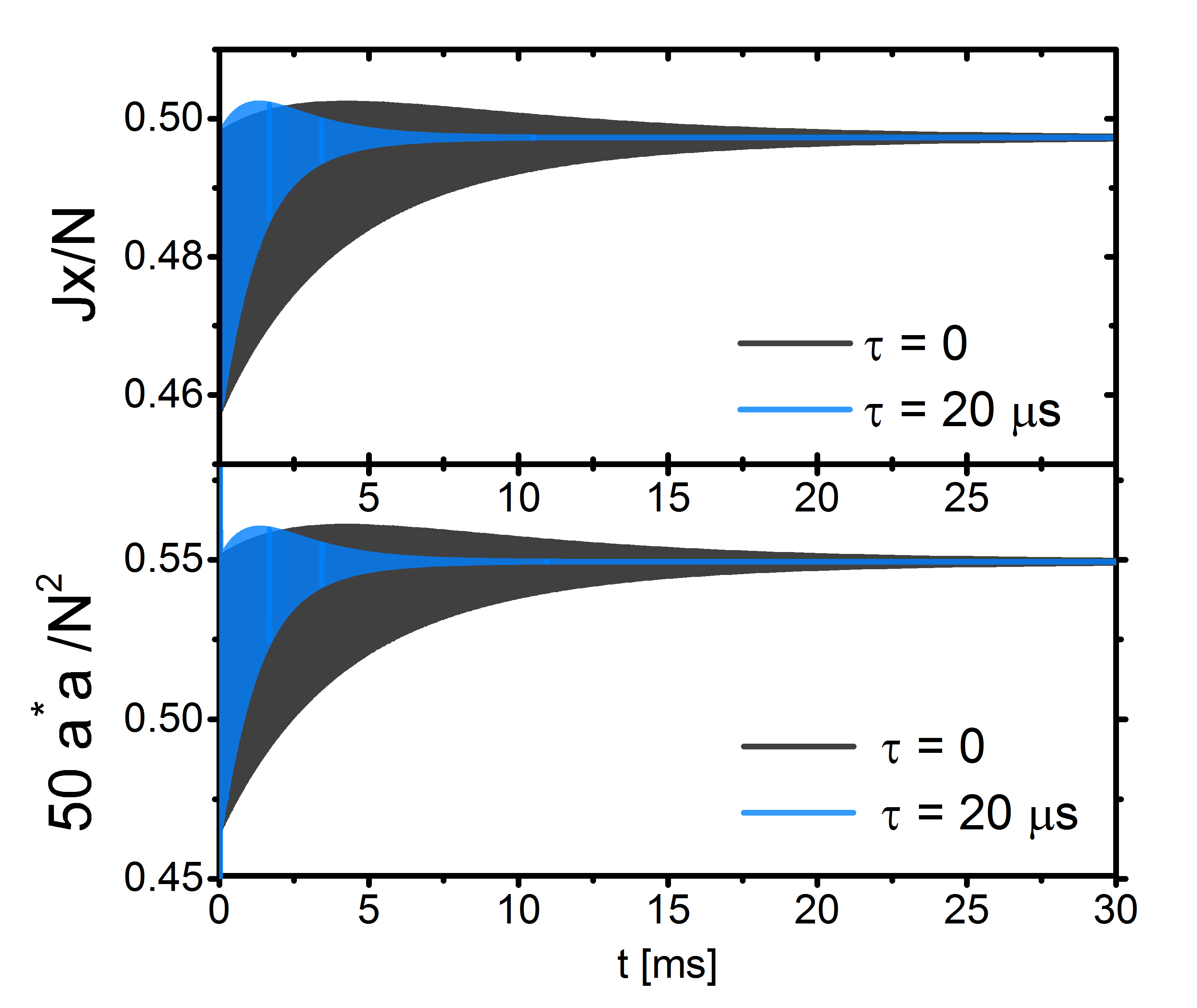}\includegraphics[width=0.55\columnwidth]{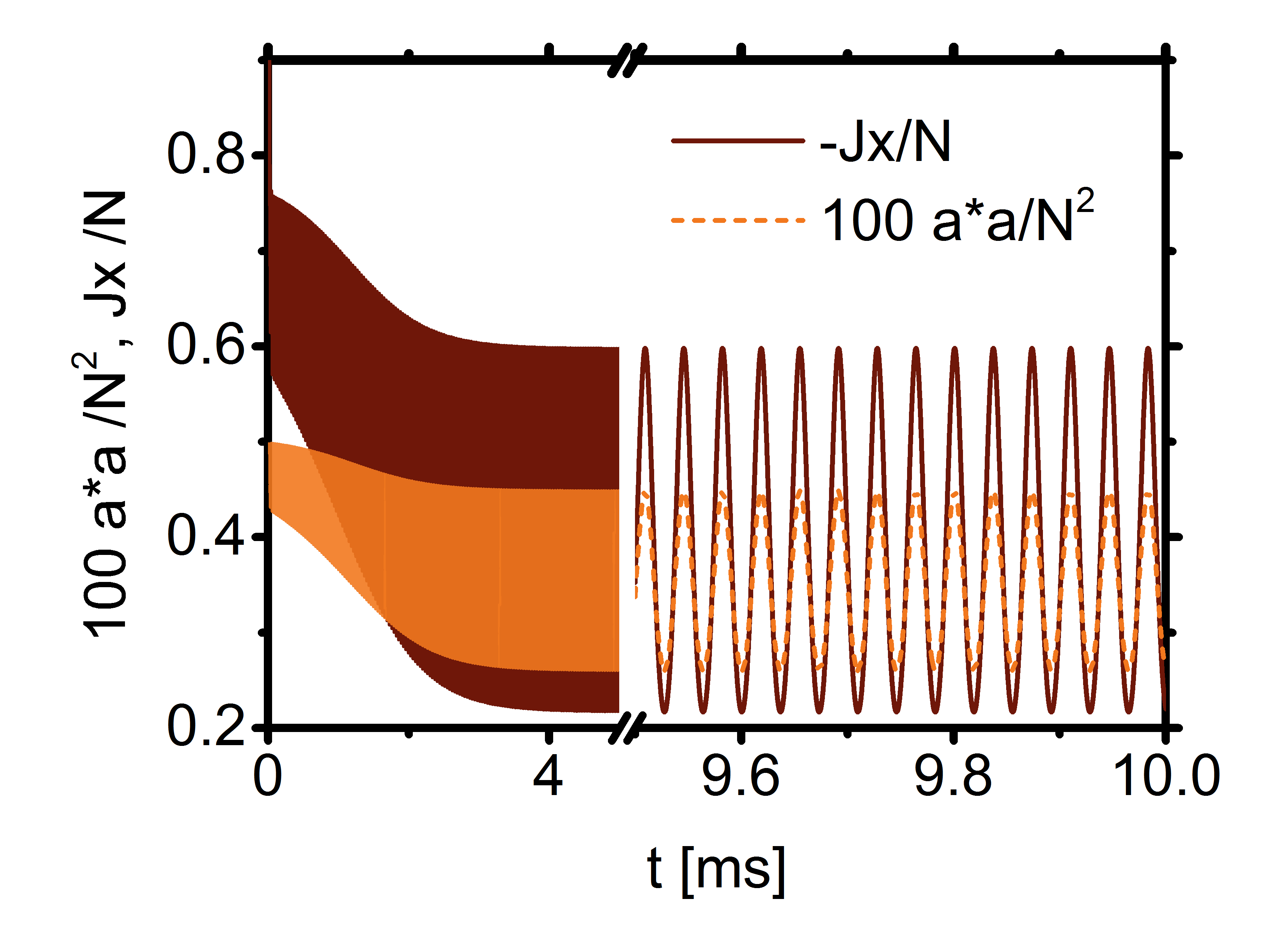}}
\caption[]{\label{figure2}
Time evolution of mode occupation $a^* a$  and angular momentum component $J_x$   corresponding to stable fixed point P2 (left) and stable limit cycle P1 (right) in the phase diagram Fig. 1.
}
\end{figure}

Note, that the $J_z$ component is determined by the conservation of
angular momentum.
Using the ansatz $\delta\mathbf{v} = \delta\mathbf{v} \, e^{\Lambda t}$, 
we obtain the characteristic equation
\begin{equation}\label{eq1:stabilitaetsbedingung}
\det\rb{\Lambda  \mathbf{1} - \mathbf{B} - \mathbf{A}  e^{-\Lambda \tau}} = 0 \,.
\end{equation}
For $\tau \neq 0$ this  transcendental equation  has an infinite set of solutions for the
eigenvalues $\Lambda \in \mathbb{C}$.
The fixed point $\mathbf{v}^{\,0}$ is stable if the real parts of all solutions
$\Lambda$
are negative, in which case the fluctuations
decay to zero for $t \to \infty$.

\textit{Phase diagrams. ---}
We obtain the phase diagram of our model in the $\omega$--$g_0$--plane (Fig. 1) from the numerical solution of Eq.~\eqref{eq1:stabilitaetsbedingung} for $\Lambda$.
First, in the left part of the phase diagram (for $g_0\le g_c$)  we recover the usual normal phase,  where  the boson occupation is zero and as a consequence, the feedback scheme Eq. (\ref{eq1:g_modulation_noninvasiv})  remains without effect.
In contrast, for $g_0> g_c$ and positive $\tau$, the superradiant phase splits up into an infinite sequence of tongue-like areas that alternate between zones with stable, superradiant fixed points (F),  and  {\em limit cycles} (L) with periodically oscillating system observables. We will devote the rest of this Letter to analysing and interpreting this rather surprising effect.

Fig. 2 displays the two markedly different types of time evolution in the superradiant regime: in the fixed point zones (F),
the only effect of the Pyragas scheme is to speed up the convergence
of the spin-components 
and the mean boson occupation $a^*a$
towards their fixed point values.
This has to be contrasted with the limit-cycle zones (L), where the fixed point is unstable, and the observables end up oscillating with a single frequency.

\begin{figure}[t]
\centerline{\includegraphics[width=\columnwidth]{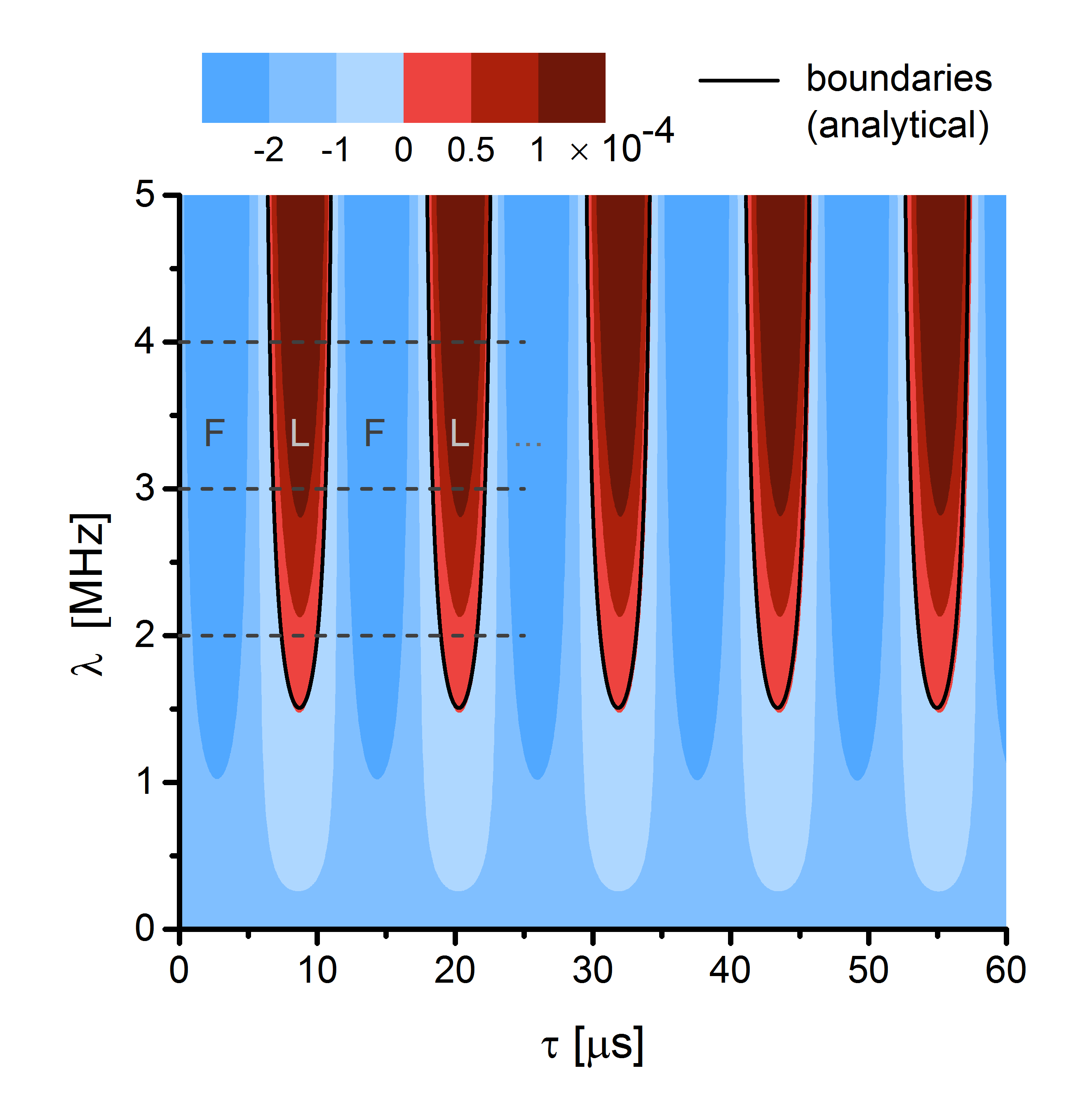}}
\caption[]{\label{figure3}
Phase diagram with sequence of stable fixed point (F) and limit cycle (L) zones in the $\lambda$ (coupling strength) vs. $\tau$ (delay time) plane. The black lines represent zone boundaries derived from the single trancendental Eq. \eqref{eq1:stabilitaetsbedingung}. Dashed lines indicate cross-sections shown in Fig. 4. Color represents the largest real part of the eigenvalues.
Parameters:$\omega=10$ MHz, $g_0 = 1.5$ MHz, $\omega_0 = 0.05$ MHz, $\kappa = 8.1$ MHz.
}
\end{figure}

\begin{figure}[t]
\centerline{\includegraphics[width=\columnwidth]{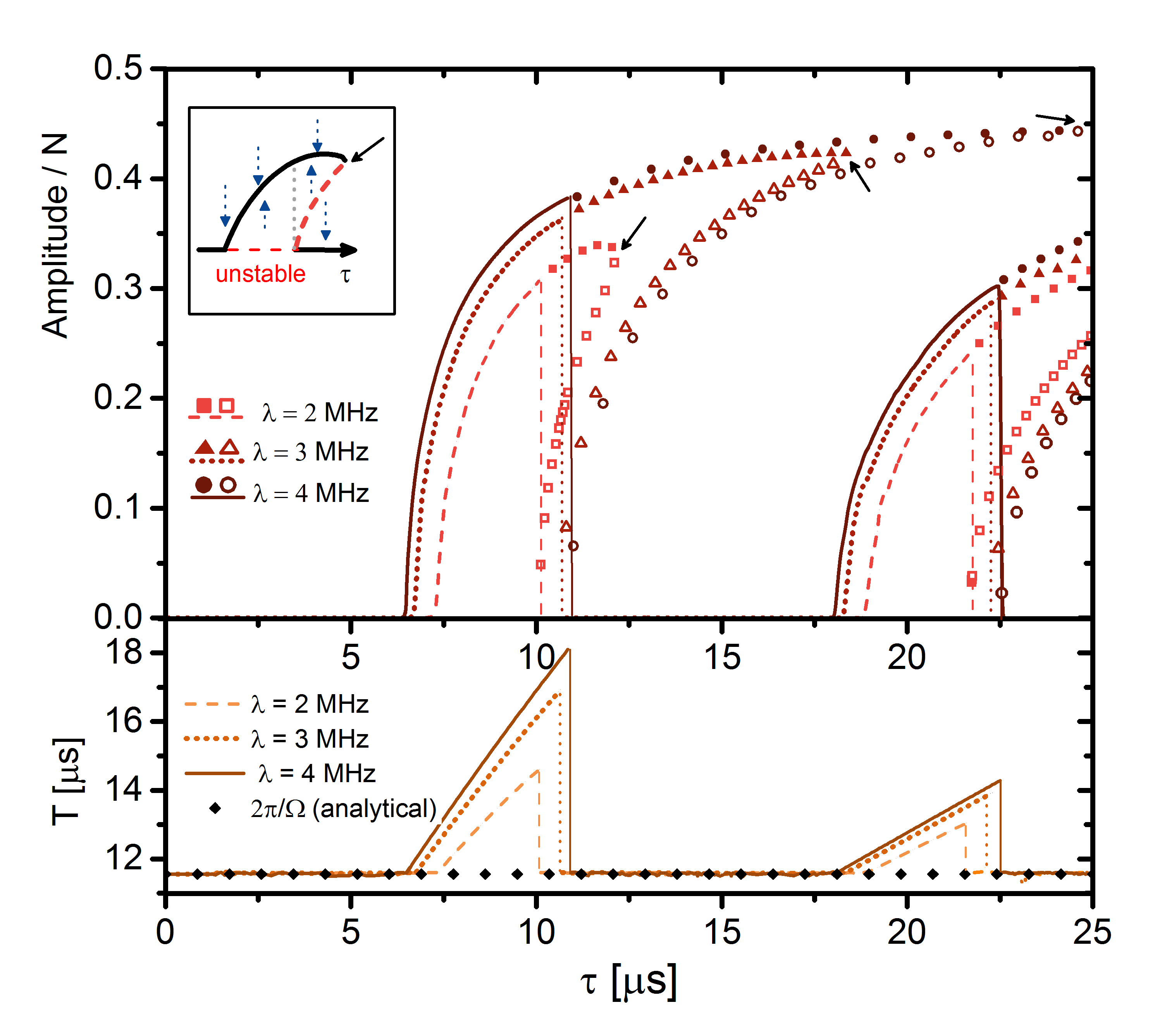}}
\caption[]{\label{figure4}
Bifurcation scenario for limit cycle amplitudes (upper)  and periods $T$ (lower) of $J_z$ as a function of delay time $\tau$ along the cross-sections (dashed lines at fixed $\lambda$) in Fig. 3. The filled symbols in the upper part describe the continuation of the limit cycle for initial value above the unstable limit cycle (unfilled symbols). The inset sketches the appearing saddle-node bifurcation of limit cycles and the dotted arrows show the direction of the phase flow for fixed tau.
}
\end{figure}

\textit{Analysis of zone boundaries. ---}
We obtain a simplified transcendental stability equation from Eq. \eqref{eq1:stabilitaetsbedingung} in the limit of very small level splitting
$\omega_0\ll \omega,g_0$, which describes the ultra-strong coupling limit of the Dicke model \cite{ABEB12} and corresponds to a feedback-controlled displaced harmonic oscillator.
In this case, the angular momenta deviations $\delta J_{x,y}$
decouple from the field deviations $\delta x,\delta y$ and describe periodic oscillations with the frequency
\begin{equation}\label{eq1:frequency}
\Omega = \frac{4 g_0^2 \omega}{N(\kappa^2+\omega^2)}.
\end{equation}
Using Eq. \eqref{eq1:stabilitaetsbedingung} we  derive an equation for $\delta \ddot{x}$ just by inserting the equations into each other.
As a result, we obtain
\begin{equation}\label{eq1:stability_end_condition_w0_is_0}
\tan(\Omega \tau) = \frac{C_2 C_3 \pm C_1 \sqrt{-C_1^2 + C_2^2 + C_3^2}}{C_1^2-C_2^2},
\end{equation}
with
$C_1 = 2 \kappa \Omega + \frac{\lambda \kappa}{g_0 \omega} \Omega^2$ ,
$C_2 = 4 g_0 \lambda$,
$C_3 = \frac{\lambda \kappa}{g_0 \omega} \Omega^2$,
which leads to the roots of the Eq. \eqref{eq1:stabilitaetsbedingung} with a vanishing real part i.e., $\Lambda = \pm i \Omega$.
Parameter configurations satisfying this equation mark the 
boundary
between stable (F) and unstable (L) fixed points which is included in Fig. 1 and matches the
numerically determined boundaries very well.

This analysis also allows us to elucidate the role of the delay time $\tau$ in the control scheme: to obtain real-valued results for the time delay $\tau$, the root in Eq. \eqref{eq1:stability_end_condition_w0_is_0} has to be positive. This condition is only satisfied if the feedback coupling $\lambda$
is larger than
some critical value $\lambda_l$, which we determine from the vanishing of  the root in Eq. \eqref{eq1:stability_end_condition_w0_is_0}. We corroborate these findings
by plotting the largest real part of the  eigenvalue numerically determined from Eq. \eqref{eq1:stabilitaetsbedingung} in the
$(\lambda,\tau)$-plane for fixed $\omega$ and $g_0$ values, see Fig. 3.
We recognize tongue-like zones switching between stable fixed-point and limit cycle (L) zones upon modification of the time delay $\tau$, and furthermore the existence of
a critical feedback strength $\lambda_l$ for entering in the (L) zones.

\textit{Limit cycle properties. ---}
Finally, we discuss the delay time $\tau$ and its role as a control parameter. The alternations between (F) and (L) zones in the superradiant phase in fact constitute an infinite sequence of super- and subcritical Hopf bifurcations of the stationary state generating stable and unstable limit cycles, respectively.
Solving the equations of motion \eqref{eq1:open_dicke_semiclass_eq_alternativ} for parameter values along the dashed lines in Fig. 3, we find
that the amplitude and period $T$ of the limit cycles depend upon tau, as depicted in Fig. 4 for the $J_z$ amplitude.
First, we recognize that for initial conditions close to the fixed point both the amplitude and the period show the same
Hopf bifurcation scenario (connected lines), with maxima, which mark the end of the L-zone,  appearing
for certain values of $\tau$.
As a particularly striking feature, we observe a drastic collapse
of the limit cycle (vertical lines) for
values of $\lambda > \lambda_l$ and the birth of an unstable limit cycle (disconnected unfilled symbols, shown only in the upper part of Fig. 4) when
the time delay $\tau$ reaches the end of the (L) zone.
Our numerics show, however, that this collapse occurs as a jump discontinuity.
Furthermore, a stable limit cycle still exists behind the (L) zone (disconnected filled symbols) as a continuation of the previous one but, because of bistability with the stable fixed point, it can only be reached if the initial amplitude lies above the amplitude of the unstable limit cycle, which marks the boundary between the basins of attraction of the fixed point and limit cycle attractors. The branches of the stable and the unstable limit cycles merge in a saddle-node-bifurcation (arrows in Fig. 4). The inset shows schematically this bifurcation, the dotted arrows point to the stable solution (black solid line) the system will take for different initial conditions.
As a consequence, the mean number of photons emitted from the system
oscillates with a fixed frequency
that can be externally controlled.
In addition, the theoretical prediction for the oscillating frequency $\Omega$ from the linear stability analysis of the fixed point Eq. \eqref{eq1:frequency} matches well with the damped oscillation period 
in the F-region, see Fig. 4.

We emphasize that the time dependence of $g(t\to \infty)$, Eq. \eqref{eq1:g_modulation_noninvasiv}, does not disappear in the L-regions in contrast to the F-regions, leading to the phase diagram discussed above.
Our feedback scheme here  switches
between non-invasive to invasive behavior by crossing the boundaries
within the phase diagram.
Our results also demonstrate that the Pyragas form in Eq. \eqref{eq1:g_modulation_noninvasiv}
is essential to create a new stable phase.
In contrast, for the direct feedback scheme $ g(t) = g_0 + \lambda \avg{\hat{a}^\dag \hat{a}}(t - \tau)$, depending on parameter values the occupation of the optical mode diverges and the control does not work well,  or the time delay does not seriously modify the phase diagram at all (not shown here).

Finally, we also checked that the (experimentally less practical)
Pyragas feedback for the angular momentum (instead of the photonic feedback)
also leads to the creation of a limit cycle phase in the super radiant regime, but can also not influence the stability of the normal phase. \\

We expect that our feedback scheme  can be implemented whenever semiclassical equations of motion provide an adequate description for
the quantum bifurcation type phase transitions that govern models with collective degrees of freedom, such as the  Dicke or the Lipkin-Meshkov-Glick model \cite{LMG-lipkin1965validity}.
An open and challenging problem is the  implementation of time-delayed feedback control for quantum critical systems beyond the mean-field level.

\textit{Acknowledgments. ---}
We thank H. Aoki, J. Lehnert, and N. Tsuji
for useful discussions. The authors gratefully acknowledge financial support from the DAAD and DFG Grants BR $1528/7-1$, $1528/8-2$, $1528/9-1$, SFB $910$, and GRK $1558$.

\end{document}